# Rapid screening of molecular beam epitaxy conditions for monoclinic $(In_xGa_{1-x})_2O_3$ alloys

Stephen Schaefer, Davi Febba, Kingsley Egbo, Glenn Teeter, Andriy Zakutayev, and Brooks Tellekamp

*National Renewable Energy Laboratory, Golden, CO, USA*

(Electronic mail: Stephen.Schaefer@nrel.gov and brooks.tellekamp@nrel.gov)

**Molecular beam epitaxy is one of the highest quality growth methods, capable of achieving theoretical material property limits and unprecedented device performance. However, such ultimate quality usually comes at the cost of painstaking optimization of synthesis conditions and slow experimental iteration rates. Here we report on high-throughput molecular beam epitaxy with rapid screening of synthesis conditions using a novel cyclical growth and *in-situ* etch method. This novel approach leverages sub-oxide desorption present during molecular beam epitaxy and as such should be broadly applicable to other material systems. As a proof of concept, this method is applied to rapidly investigate the growth space for the ternary alloy $(In_xGa_{1-x})_2O_3$ on (010) oriented β-$Ga_2O_3$ substrates using *in-situ* reflection high energy electron diffraction measurements. Two distinct growth regimes are identified and analyzed using machine learning image recognition algorithms, the first stabilizing a streaky 2× surface reconstruction typical of In-catalyzed β-$Ga_2O_3$ growth, and the second exhibiting a spotty/faceted pattern typical of phase separation. Targeted growth of $(In_xGa_{1-x})_2O_3$ is performed under conditions near the boundary of the two regimes resulting in a 980 nm thick epitaxial layer with In mole fraction up to 5.6%. The cyclical growth/etch method retains the ~1 nm surface roughness of the single crystal substrate, increases experimental throughput approximately 6×, and improves single crystal substrate utilization by >40×. The high-throughput MBE method enables rapid discovery of growth regimes for ultra-wide bandgap oxide alloys for power conversion devices operating with high efficiency at high voltages and temperatures, as well as optical devices such as ultraviolet photodetectors.**

## I. INTRODUCTION

Molecular beam epitaxy (MBE) has a long history of growing materials with world record structural, optical, and electrical properties ever since its inception in the late 1960s.[1] MBE is an ultrahigh vacuum growth technique in which one or more high purity source materials are thermally evaporated to deposit crystalline films on a suitable substrate. As a result of the extremely low ($< 10^{-9}$ torr typical) base pressure in the MBE chamber, the mean free path between gas molecules is much larger than the dimensions of the chamber and the source material fluxes are transported ballistically without scattering.[2-4] The ultrahigh vacuum achieved in MBE enables extremely low impurity incorporation rates. Impurity concentrations on the order of $10^{13}$-$10^{14}$ $cm^{-3}$ have been demonstrated in MBE-grown GaAs,[5] enabling record-high electron mobilities in excess of $10^7$ $cm^2/V\cdot s$ in GaAs/AlGaAs heterostructures.[6-8] MBE is capable of growing atomically smooth and abrupt interfaces necessary for quantum-confined structures such as quantum cascade lasers[9] or high electron mobility transistors.[10] *In-situ* characterization techniques such as reflection high energy electron diffraction (RHEED), spectroscopic ellipsometry, laser reflectometry, and mass spectroscopy provide real-time information about crystal growth and material properties. The flexibility afforded by the elemental sources, as well as gas-phase plasma sources, makes MBE uniquely suited for the growth of novel alloys and heterostructures. However, the low growth rates and throughput typical of MBE growth remains a major drawback for high-throughput experimentation. The traditional "one substrate, one experiment" MBE growth paradigm is both time-consuming and expensive, involving many iterations of growth, characterization, and refinement of growth conditions on single-use substrates costing up to ~$1,000 USD per square cm.

Gallium oxide ($Ga_2O_3$) is an emerging ultra-wide bandgap semiconductor material that has attracted attention for its potential to outperform existing SiC and GaN based devices operating at high breakdown voltages and high temperature. The thermodynamically stable phase at room temperature and pressure is the monoclinic β-phase with symmetry *C2/m*. β-$Ga_2O_3$ has a direct bandgap energy of 4.76 eV[11] and critical field as high as 8 MV/cm.[12,13] Additionally, β-$Ga_2O_3$ exhibits controllable n-type doping with high ionization efficiency,[14-17] robust mechanical properties, and high-quality large-area growth from melt,[13,14,16,18] making it a strong candidate material for next-generation power electronic devices operating at high voltages. β-$Ga_2O_3$ based materials are particularly attractive for power conversion devices operating in extreme environments and high temperatures due to their inherent stability against degradation due to oxidation. Ultra-wide bandgap $Ga_2O_3$ based materials find further application in energy and sustainability as HEMTs,[19-23] solar-blind ultraviolet photodetectors,[24-27] and ferromagnets.[28]

Isovalent alloying of In and Al in β-$Ga_2O_3$ provides the ability to engineer bandgap energy and strain of the material. First-principles calculations indicate a range of bandgap energies from 7.2-7.5 eV[29,30] for monoclinic θ-$Al_2O_3$ to 2.7 eV[31,32] for monoclinic $In_2O_3$. The corresponding lattice mismatch to β-$Ga_2O_3$ ranges from about 4% for θ-$Al_2O_3$[33,34] to 10% for monoclinic $In_2O_3$.[31,33] However, epitaxial growth of the complementary ternary alloy $(In_xGa_{1-x})_2O_3$ has proven more challenging than β-$Ga_2O_3$ or even $(Al_xGa_{1-x})_2O_3$., due to competing structural phases, indium and indium oxide volatility, and multi-step reaction processes. An additional complicating feature of β-$Ga_2O_3$ MBE growth is etching under Ga-rich conditions, where $Ga_2O$ readily desorbs at typical growth temperatures resulting in negative growth rate. However, the same suboxide desorption during Ga etching of $Ga_2O_3$ and $(In_xGa_{1-x})_2O_3$ can be leveraged for cyclic *in-situ* growth studies by etching away the grown film to recover the initial β-$Ga_2O_3$ substrate surface while monitored by RHEED. Unlike conventional chemical wet etching and annealing, the substrate surface is recovered *in-situ* without requiring time-consuming vacuum transfer, cleaning, and annealing steps.

Here we show cyclical $(In_xGa_{1-x})_2O_3$ growth and Ga flux etch-back of the grown film down to the β-$Ga_2O_3$ substrate to rapidly screen the multi-variable conditions of the plasma-assisted MBE $(In_xGa_{1-x})_2O_3$ growth space. RHEED measurements of the specular streak width enable a highly repeatable growth/etch process that recovers the initial β-$Ga_2O_3$ surface with < 1.6 nm surface roughness after 46 etch-back cycles. Two growth regimes are identified using machine learning image recognition methods based on RHEED patterns, and the patterns are analyzed for similarity to reference images representative of β-$Ga_2O_3$ and bixbyite $In_2O_3$. A targeted $(In_xGa_{1-x})_2O_3$ growth is carried out at conditions near the growth boundaries where enhanced In incorporation in the monoclinic phase is anticipated. The targeted growth is characterized *ex-situ* by X-ray diffraction (XRD), X-ray photoelectron spectroscopy (XPS), Rutherford back-scattering spectrometry (RBS), and atomic force microscopy (AFM). The results lay the foundation for growth of high In content β-$(In_xGa_{1-x})_2O_3$ and band structure engineering in gallium oxide heterostructures. The cyclical growth/etch method could be extended to alloy growths in oxide and other materials exhibiting incongruent evaporation of the component gas species[34,35] such as $SnO_2$, $GeO_2$, $In_2Se_3$, and $Ga_2Se_3$.

## II. METHODS AND APPROACH

### *1. MBE and RHEED*

β-$(In_xGa_{1-x})_2O_3$ films were grown epitaxially on (010) oriented Fe doped 5mm × 5mm β-$Ga_2O_3$ wafers provided by Novel Crystal Technology. Wafers were cleaned by a sequential acetone/methanol/2-propanol solvent clean followed by two sequential 10-minute cleans in 4:1 $H_2SO_4$:$H_2O_2$ followed by a DI water rinse. The substrates were then indium bonded to 2" diameter Si carrier wafers using 6N5 purity indium and loaded into Mo platens. Plasma-assisted MBE was performed in a Riber Compact 21 T system. 7N Ga was sourced from a SUMO style effusion cell while 6N5 In was sourced from a conventional dual-filament effusion cell. High purity dry oxygen (SAES purifier, <1 part per billion $H_2O$ by volume) was supplied through a 13.56 MHz radio-frequency oxygen plasma source (Veeco UNI-Bulb) operating at 250 W at a standard flow rate of 3.0 SCCM. Substrates were outgassed in an introductory vacuum chamber at 150 °C for at least 120 minutes before transferring to an intermediate buffer chamber. Ga and In fluxes were measured using a Bayard-Alpert style retractable ionization gauge before each growth. The growth temperature of the β-$Ga_2O_3$ wafers was calibrated to the non-contact substrate thermocouple temperature using a UV band edge thermometry (UV-BandiT) system provided by k-Space Associates, Inc.

Films were analyzed *in-situ* using a differentially pumped RHEED system provided by Staib Instruments and operated at 20 kV beam voltage and 10 μA emission current. The differential pumping stage provides the ability to operate RHEED during oxygen plasma-assisted growth. RHEED patterns were collected by a CCD camera at a standard 83 ms exposure time. Patterns were recorded and analyzed in real time using k-Space Associates kSA 400 software. kSA 400 was used to track peak intensity, in-plane *d*-spacing, and full width half-maximum (FWHM) as a function of time. Discrete RHEED images were also periodically acquired. The RHEED pixel *d*-spacing was calibrated to the streak spacing in the known diffraction pattern from the <001> azimuth of β-$Ga_2O_3$ with *c*-plane lattice constant of 5.7981 Å without adjusting for thermal expansion.[33]

The diffraction pattern recorded on the RHEED screen is a reciprocal space image of the (nominally) two-dimensional surface electron density. In the ideal case of an atomically smooth surface with crystal domain sizes greater than the RHEED beam coherence length (typically 100-200 nm),[36] the reciprocal lattice of a crystalline material consists of rods perpendicular to the surface with spacing inversely proportional to the lattice periodicity. The intersection of the Ewald sphere formed by elastically scattered electrons with the reciprocal lattice rods results in a pattern of diffraction spots on the RHEED screen. In practice, the energy dispersion of the diffracted electron beam and sample non-idealities, e.g. finite domain size and atomic-scale surface roughness, broaden the reciprocal rods.[36,37] The resulting RHEED pattern consists of streaks, where the streak width (FWHM) is inversely proportional to the domain size or the diffraction coherence length. The diffraction streaks are indexed with reference to the specular reflection (00) along the <001> azimuth. First-order streaks correspond to the lattice periodicity along the [001] direction, while half order streaks $\left(0\frac{1}{2}\right)$, $\left(0\frac{\bar{1}}{2}\right)$, etc. correspond to a reconstructed surface with twice the lattice periodicity, e.g. dimerization of dangling bonds. This is referred to as a "2×" surface reconstruction.[36-38]

### 2. *Cyclical β-$(In_xGa_{1-x})_2O_3$ growth and etch*

A method of cyclical β-$(In_xGa_{1-x})_2O_3$ growth and Ga etch was developed to rapidly investigate the MBE growth space using RHEED. The method is illustrated schematically in Fig. 1a. The RHEED beam is first oriented perpendicular to the <001> azimuth of the (010) oriented β-$Ga_2O_3$ wafer at a thermocouple temperature of 800 °C (wafer temperature of 735-740 °C). Ga flux of $4.0\times10^{-7}$ torr beam equivalent pressure (BEP) is supplied for 15 minutes in the absence of oxygen plasma as an initial etch and *in-situ* cleaning step. The wafer is then exposed to the oxygen plasma for 15-30 minutes in the absence of Ga flux to heal crystalline imperfections from wafer cutting and polishing.[39] This oxygen annealing step results in a slight reduction in the specular spot (00) FWHM to a typical value of 8.5 – 9 pixels, indicating a smooth and well-ordered surface. $(In_xGa_{1-x})_2O_3$ growth proceeds by exposing the wafer to the oxygen plasma and Ga and In fluxes. Ga fluxes range from $5\times10^{-8}$ to $2.0\times10^{-7}$ torr while

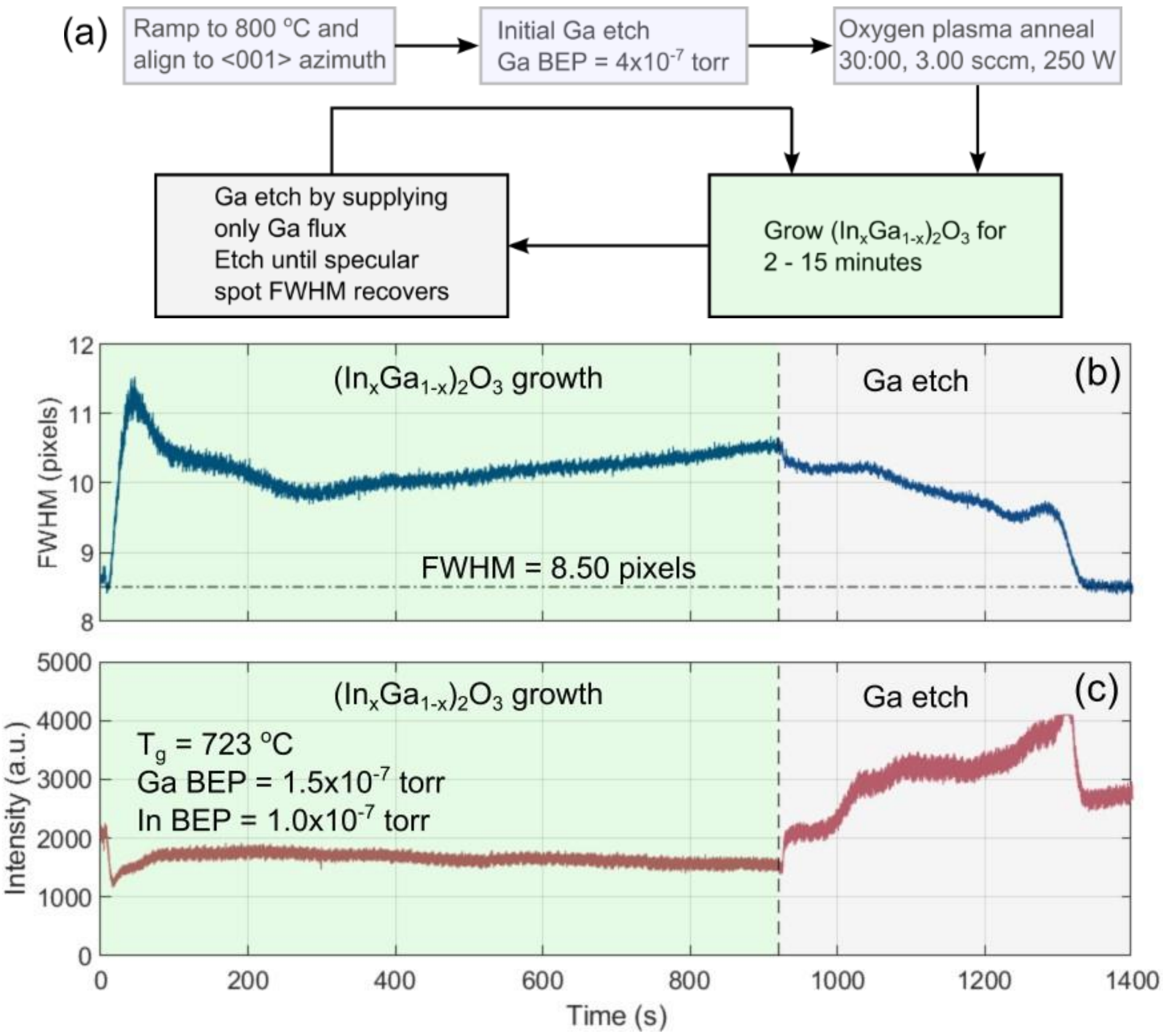

FIG. 1. Cyclical $(InGa)_2O_3$ MBE growth and Ga etch. (a) Flowchart for growth/etch-back experiments, including initial substrate preparation. (b) Specular spot full-width half-maximum (middle, blue) and (c) intensity (bottom, red) vs time for a complete growth/etch cycle. The FWHM increases upon initiation of $(InGa)_2O_3$ growth and decreases during the Ga etch until recovering to its initial value, indicating complete removal of the $(InGa)_2O_3$ epilayer.

In fluxes range from $5\times10^{-8}$ to $1.0\times10^{-6}$ torr. The RHEED pattern specular spot (00) FWHM and peak intensity are tracked in real time, as well as the spacing between diffraction streaks. $(In_xGa_{1-x})_2O_3$ growth is terminated when the RHEED pattern reaches a steady state, as short as 2 minutes, or when 15 minutes have elapsed, whichever condition is satisfied sooner. The wafer is then exposed to ~$4\times10^{-7}$ torr Ga at a thermocouple temperature of 800 °C to etch back the $(In_xGa_{1-x})_2O_3$ layer. Ga etching proceeds until the (00) FWHM recovers to its initial minimum value, indicating complete removal of the $(In_xGa_{1-x})_2O_3$ layer and etch back into the β-$Ga_2O_3$ wafer. The evolution of the (00) spot FWHM and peak intensity are shown in Fig. 1b and Fig. 1c for a complete growth/etch cycle. As shown in the Ga etch right-hand pane, the completion of film etching is accompanied by a characteristic drop in FWHM which is repeatable and easily recognized. There is no fundamental limitation to the number of substrate re-use cycles apart from possible over-etching into the finite thickness (~500 um) of the $Ga_2O_3$ wafer.

### 3. *Machine learning of RHEED patterns*

A total of 109 $(In_xGa_{1-x})_2O_3$ growth experiments were conducted using the cyclical growth/etch method. Manual comparison of RHEED patterns with reference images is time-consuming and prone to subjective bias, which can lead to inconsistent or inaccurate evaluations. An automated, objective method for comparing these patterns is therefore desirable to enhance the accuracy and efficiency of the analysis, and to provide more reliable insights into material characteristics. Several automated approaches could be employed for this task, including deploying a deep learning classifier or logistic regression. However, these approaches often require large amounts of labelled training data or rely on handcrafted feature engineering. In this work the dataset of RHEED images underwent a preprocessing routine followed by an automated similarity assessment using the Learned Perceptual Image Patch Similarity (LPIPS) metric,[40] carried out using the PyTorch framework. Given the dataset size, the LPIPS metric stands out as it leverages a pre-trained convolutional neural network (AlexNet in this work), alleviating the need for additional training data and thereby presenting a suitable solution for this analysis. Moreover, LPIPS offers a significant advantage over traditional metrics like cosine similarity, as it is designed to align closely with human visual perception, making it particularly effective in identifying subtle yet perceptually significant differences in image patterns.

The preprocessing, using the scikit-image library, started with the conversion of grayscale images to RGB, which is a requirement for the subsequent analysis with LPIPS. The images were then cropped to retain only the upper half, eliminating diffuse scattering at high diffraction angles while retaining the surface reconstruction information contained within the first Laue circle. The Contrast Limited Adaptive Histogram Equalization (CLAHE) correction was applied via the *OpenCV* library with a clip limit of 10.0 and a tile grid size of 8×8 (default parameter) to improve the contrast in each image, enhancing the visibility of patterns. Lastly, the pixel values were normalized to a range of [−1,1], in alignment with the requirements of the LPIPS metric.

### 4. *Ex-situ analysis of β-$(In_xGa_{1-x})_2O_3$ layers*

A targeted $(In_xGa_{1-x})_2O_3$ growth was characterized *ex-situ* by XRD, XPS, and RBS. Grown and etched samples were also characterized by AFM. XRD was performed using a Rigaku SmartLab diffractometer with Cu-Kα radiation monochromated by a 2-bounce Ge (220) crystal, or a Panalytical MRD Pro diffractometer using Cu-Kα radiation and a hybrid monochromator (Göbel mirror and a 4-bounce Ge (400) crystal) on the incident beam. XPS measurements were performed using Physical Electronics

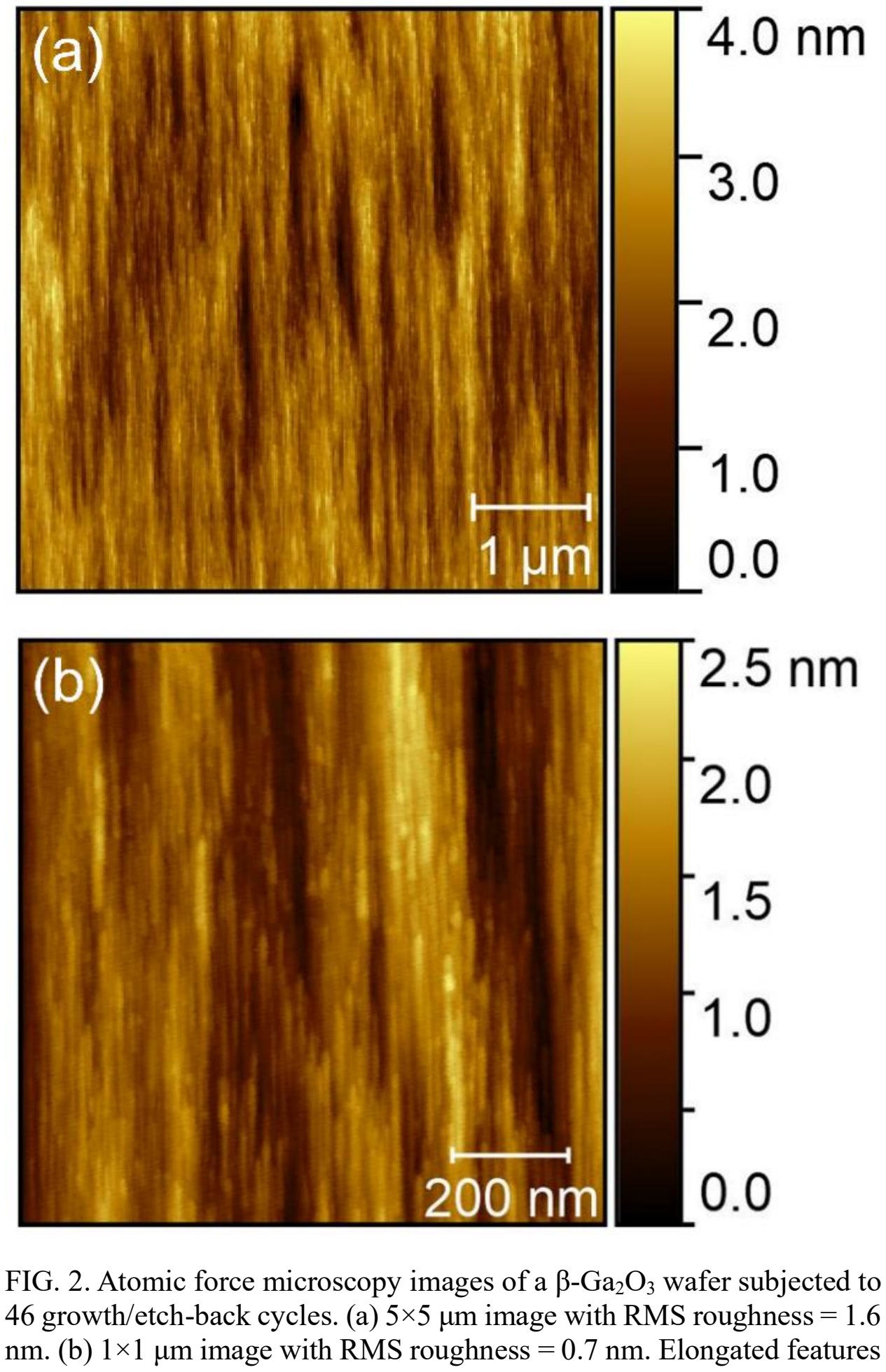

FIG. 2. Atomic force microscopy images of a β-$Ga_2O_3$ wafer subjected to 46 growth/etch-back cycles. (a) 5×5 µm image with RMS roughness = 1.6 nm. (b) 1×1 µm image with RMS roughness = 0.7 nm. Elongated features are aligned along the [001] direction and are commonly observed after Ga etching of (010) oriented $Ga_2O_3$ [41].

Phi VersaProbe III with monochromatic Al-kα excitation (hν = 1486.7 eV) at 69 eV pass energy. XPS depth profiles were performed using an argon gas-cluster ion beam (GCIB) source operated at 20 keV incident energy, with typical cluster size ~2500 argon atoms. RBS measurements were performed by Evans Analytical Group using a 2.275 MeV $He^{2+}$ ion beam and detector angles of 160° (normal angle) and 100° (grazing angle). The backscattered spectra are fit by a multilayer theoretical model to determine the Ga and In atomic concentrations as a function of depth. AFM images are acquired in tapping mode using a Veeco DI D3100 AFM equipped with a Nanoscope V controller. Fig. 2 shows AFM images of a β-$Ga_2O_3$ wafer subjected to 46 cycles of growth and etch-back. The RMS roughness is 1.6 nm over a 5×5 μm area and 0.7 nm over a 1×1 μm area, confirming the etch-back process recovers an atomically smooth growth surface after each $(In_xGa_{1-x})_2O_3$ growth.

## III. RESULTS AND DISCUSSION

### *1. RHEED surface reconstruction growth map*

$(In_xGa_{1-x})_2O_3$ growths were performed at BandiT calibrated growth temperatures ranging from 592-771 °C, Ga fluxes ranging from 0.5-2.0×$10^{-7}$ torr, and In fluxes ranging from 0-1.0×$10^{-6}$ torr at a fixed oxygen plasma power of 250 W and 3.00 SCCM flow. Growth durations ranged from 2 to 15 minutes ending with the stabilization of one of two distinct RHEED patterns shown in Fig. 3. The terminal RHEED patterns are a streaky 2× surface reconstruction typical of the <001> azimuth of β-$Ga_2O_3$ (Fig. 3a) or a spotty/faceted pattern attributed to the formation of bixbyite $In_2O_3$ on the growth surface (Fig. 3b). The evolution of the principal streaks into the broad spots in Fig. 3b indicates significant surface roughening and transmission through a distinct crystal phase on the growth surface, while the appearance of three higher-order spots in between the principal (01) spots suggests the formation of facets with approximately 4× the periodicity of the crystal lattice. Additionally, the *d*-spacing of the principal spots/streaks increases from 5.8 Å for ungrown β-$Ga_2O_3$ to 7.1 Å for the spotty/faceted pattern in Fig. 3b. These diffraction spots are consistent with diffraction from the bixbyite (110) plane with RHEED pattern periodicity $a/\sqrt{2}$ = 7.155 Å.[42] The possible orientation of bixbyite $In_2O_3$ on (010) β-$Ga_2O_3$ is shown in the Supplemental Information. The typical time to stabilize the spotty/faceted pattern in Fig. 3b is 2-3 minutes. In contrast the streaky 2× surface reconstruction is stable to 15 minutes or longer, with some degradation in (00) and (01) FWHM over time observed at transitional growth conditions (discussed below). Crucially, however, the lattice *d*-spacing remains stable and the periodicity of the 2× reconstruction does not change.

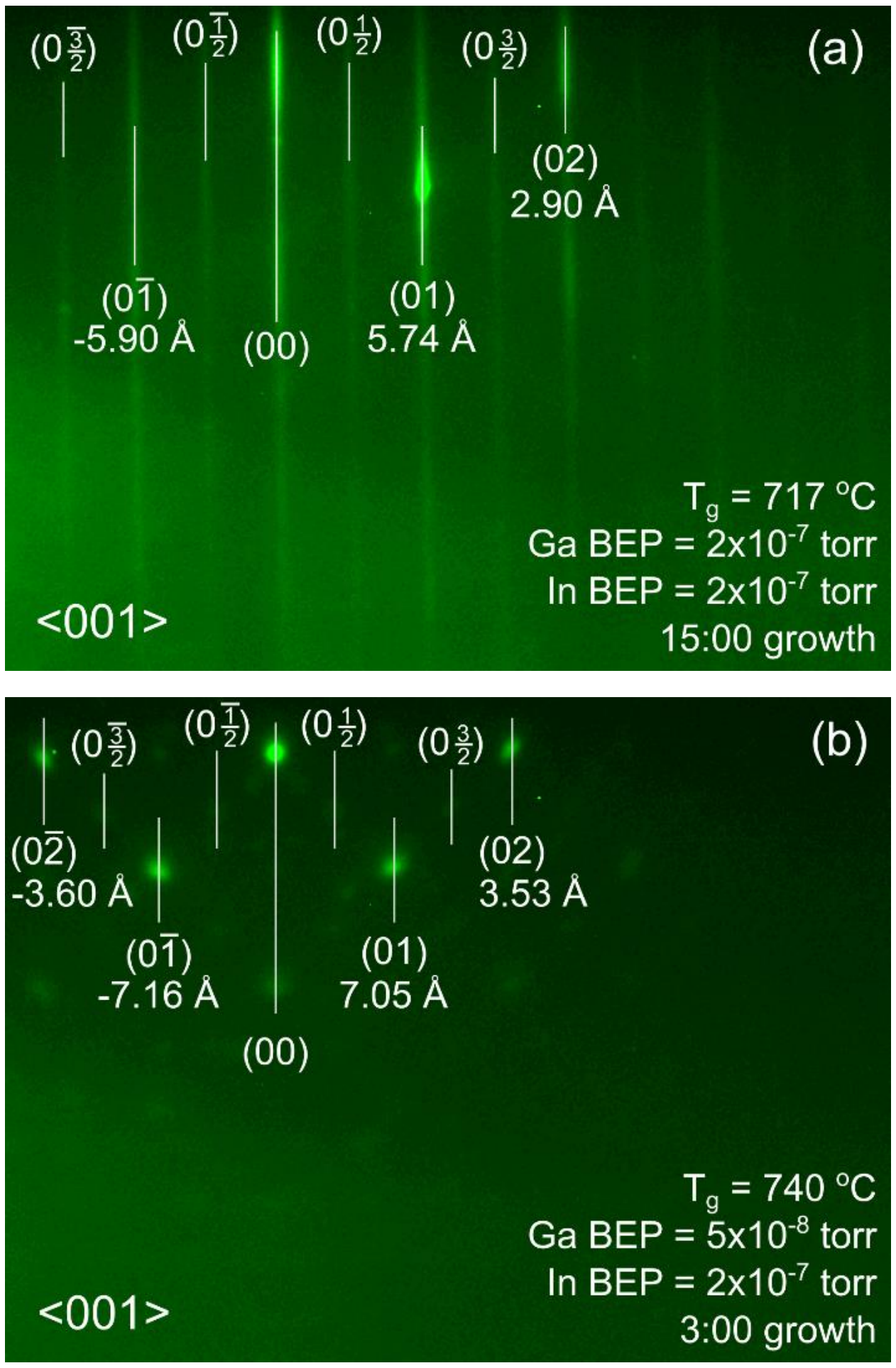

FIG. 3. Terminal RHEED patterns of (a) streaky 2x surface reconstruction typical of $Ga_2O_3$ growth, and (b) spotty/faceted pattern typical of low-temperature, Ga-lean growth. Streaks are labeled with *d*-spacing relative to the (00) streak. The {01} streak *d*-spacing for the spotty/faceted pattern in (b) is attributed to diffraction from the bixbyite $In_2O_3$ (110) plane.

Fig. 4 shows the growth map of $(In_xGa_{1-x})_2O_3$ alloys as a function growth temperature, Ga flux, and In flux space. This RHEED data for this growth map was collected by the cyclical growth/etch method and analyzed using a machine learning algorithm with a LPIPS metric and CLAHE pre-processing. The terminal RHEED patterns for each growth setpoint were classified as either streaky 2× (similar to Fig. 3a) or spotty/faceted (similar to Fig. 3b). The Ga and In beam equivalent pressures were converted into particle fluxes according to the procedure in the Supplemental Information. The particle fluxes and resulting metal/oxygen ratio $(Ga + In)_2O_3$ are plotted in Fig. 4 (b-d). A growth boundary between the two classes of RHEED patterns is observed, with lower growth temperatures and lower metal fluxes resulting in the spotty/faceted pattern associated with formation of bixbyite $In_2O_3$. Above approximately 725 °C growth temperature, or metal/oxygen flux ratio greater than 2/3, the bixbyite $In_2O_3$ RHEED pattern is suppressed and the growth surface exhibits a 2× reconstruction typical of monoclinic phase In-catalyzed $Ga_2O_3$ or $(In_xGa_{1-x})_2O_3$ growth. The growth boundary is strongly dependent on growth temperature and total metal/oxygen flux ratio, with relatively weak dependence on the In flux.

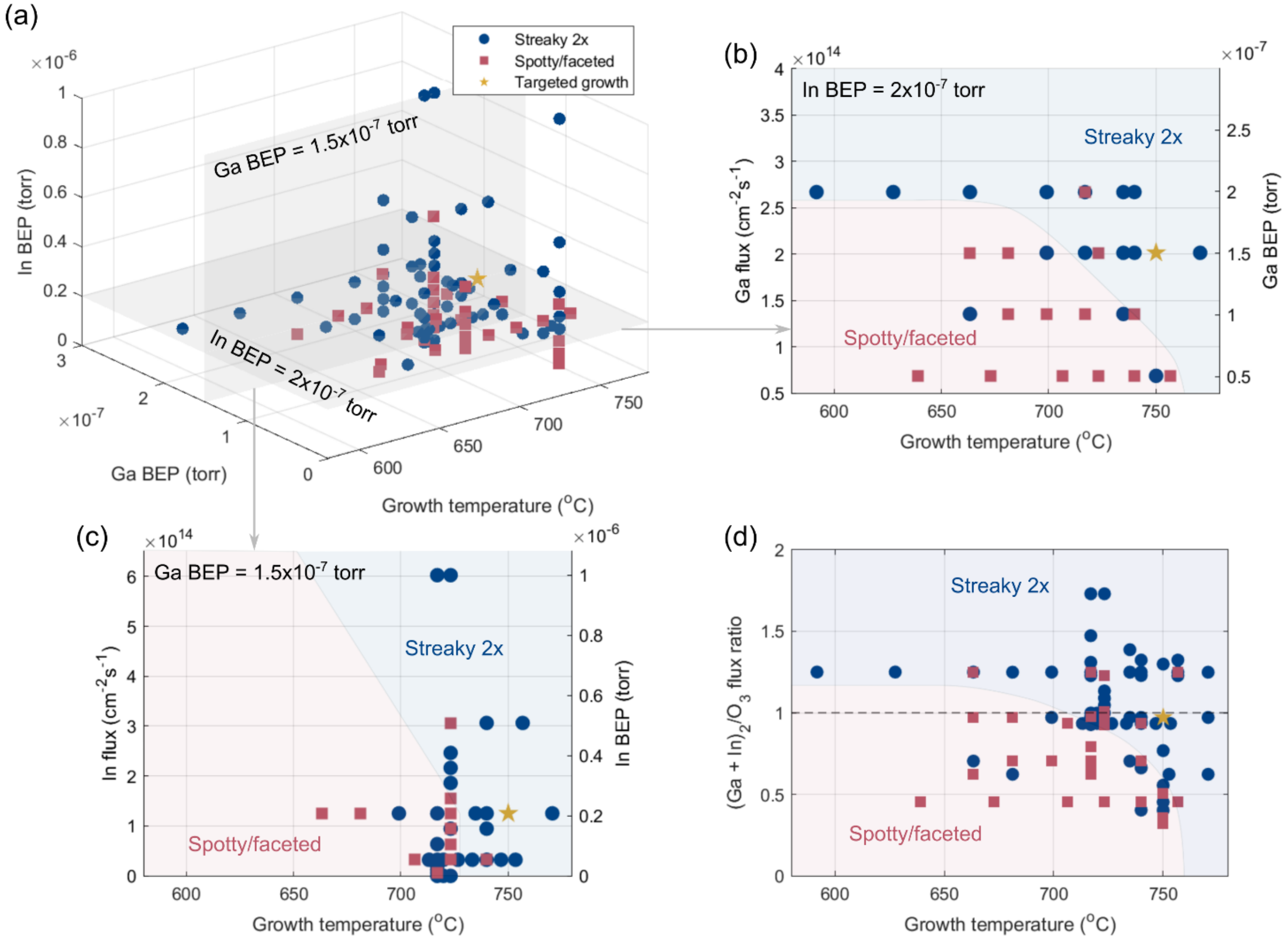

FIG. 4. Growth map of $(In_xGa_{1-x})_2O_3$ alloys obtained from cyclical growth/etch experiments with results classified by similarity analysis. (a) RHEED surface reconstruction vs Ga BEP, In BEP, and growth temperature $T_g$. Blue circles: Streaky 2× reconstruction stabilized. Red squares: Spotty/faceted reconstruction. Gold star: Targeted thick growth. (b) Reconstruction as a function of $T_g$ and Ga flux at In BEP = $2\times10^{-7}$ torr. (c) Reconstruction as a function of $T_g$ and In flux at Ga BEP = $2\times10^{-7}$ torr. (d) Reconstruction as a function of $T_g$ and stoichiometric flux ratio $(Ga + In)_2/O_3$. Shaded regions are guides to the eye.

It's notable that the $(In_xGa_{1-x})_2O_3$ alloy growth map shown in Fig. 4 was created using automated machine learning algorithm LPIPS for RHEED pattern recognition, rather than traditional manual labeling and sorting of RHEED patterns. Unlike traditional metrics such as Mean Squared Error (MSE) or Peak Signal-to-Noise Ratio (PSNR) that analyze images on a pixel level, the LPIPS metric used here operates in a feature space. In this scenario, the architecture of a pre-trained CNN known as AlexNet[43] can be employed to extract features from images. Briefly, a feature space is a representation where images are characterized by a set of descriptors, or features, that encapsulate informative aspects beyond pixel values, such as edges, textures, or shapes. Through AlexNet, LPIPS transforms the images into this feature space, facilitating a comparison that highlights more meaningful similarity between images. This approach has shown alignment with human perceptual judgments across various datasets and image distortions, providing a more perceptually aligned measure of similarity.[40]

### 2. *Chemistry of In-catalyzed $(In_xGa_{1-x})_2O_3$ growth*

Epitaxial growth of the ternary alloy $(In_xGa_{1-x})_2O_3$ has proven more challenging than $(Al_xGa_{1-x})_2O_3$ due to the unique (In,Ga) oxide growth chemistry variously referred to as metal oxide catalyzed epitaxy ("MOCATAXY")[44-46] or metal-exchange catalysis ("MEXCAT").[47,48] The addition of In flux during MBE growth of $Ga_2O_3$ results in a 2-step reaction characterized by the formation of $In_2O_3$ followed by metal exchange between Ga and In to form β-$Ga_2O_3$:[49]

$$2In(a) + 3O(a) \rightarrow In_2O_3(s) \quad (1)$$

$$2Ga(a) + In_2O_3(s) \rightarrow Ga_2O_3(s) + 2In(a) \quad (2)$$

This metal exchange reaction increases the effective oxygen reservoir at the growth surface due to the higher oxidation efficiency of In compared to Ga,[34,49] resulting in an increase in $Ga_2O_3$ growth rate. However, negligible In incorporation occurs in this In-catalyzed growth regime,[41,44-49] with the remaining adsorbed indium in reaction (2) desorbing from the growth surface. MBE growth conditions for incorporating indium in single crystal monoclinic $(In_xGa_{1-x})_2O_3$ alloys remain elusive. Mauze *et al* reported In incorporation up to approximately 24% mole fraction for MBE growth at low Ga/O flux ratio at 900 °C and 800 °C on (010) and (001) oriented $Ga_2O_3$ substrates, respectively.[45] Von Wenckstern *et al* synthesized polycrystalline $(In_xGa_{1-x})_2O_3$ across the entire composition range by pulsed laser deposition on *c*-plane sapphire.[50] Few reports of successful synthesis of thick monoclinic $(In_xGa_{1-x})_2O_3$ epilayers with non-negligible In mole fractions exist.[44,45]

An additional unique characteristic of β-$Ga_2O_3$ MBE growth is etching under Ga-rich conditions. The $Ga_2O_3$ growth rate decreases with increasing Ga/O flux ratio due to the formation of the volatile suboxide $Ga_2O$:[34,51,52]

$$4Ga(a) + Ga_2O_3(s) \rightarrow 3Ga_2O(a) \quad (3)$$

$Ga_2O$ readily desorbs at typical growth temperatures resulting in negative growth rate, or etching, under Ga/O flux ratios greater than three.[34,52] Furthermore, highly Ga-rich conditions lead to etching of $(In_xGa_{1-x})_2O_3$ as In is readily exchanged by Ga according to reaction (2) to form $Ga_2O_3$, followed by subsequent desorption of adsorbed In and $Ga_2O$ according to reaction (3).

Based on the Eq.1-3 presented above, the $(In_xGa_{1-x})_2O_3$ alloy map presented in Fig. 4 can be explained by the chemistry of In-catalyzed growth in a two-step process:

(a) In the first step, adsorbed In is preferentially oxidized to form either $In_2O_3$ or $In_2O$ on the growth surface at all growth temperatures investigated. There are three reaction regimes based on the In/oxygen flux ratio $r_{In} = \Phi_{In}/\Phi_O^{*,In}$. In the oxygen-rich regime with $r_{In} < 1$, $In_2O_3$ forms with excess adsorbed oxygen similar to Eq. (1):[53]

$$2r_{In}In(a) + 3O(a) \rightarrow r_{In}In_2O_3(s) + 3(1-r_{In})O(a) \quad (4)$$

In the moderately In-rich regime (ii) with $1 < r_{In} < 3$, excess In reacts with $In_2O_3$ to form $In_2O$, reducing the total amount of $In_2O_3$ on the surface:[41]

$$2r_{In}In(a) + 3O(a) \rightarrow In_2O_3(s) + 2(r_{In}-1)In(a), \quad (5a)$$

$$In_2O_3(s) + 2(r_{In}-1)In(a) \rightarrow \tfrac{3}{2}(r_{In}-1)In_2O(a) + \tfrac{1}{2}(3-r_{In})In_2O_3(s) \quad (5b)$$

In the highly In-rich regime with $r_{In} > 3$, only adsorbed $In_2O$ and In are present similar to Eq. (3):

$$2r_{In}In(a) + 3O(a) \rightarrow 3In_2O(a) + 2(r_{In}-3)In(a) \quad (6)$$

(b). In the second step, Ga flux reacts with both excess adsorbed O to directly form $Ga_2O_3$ and with $In_2O_3$ by means of the cation exchange reaction (2) to form the alloy $(In_xGa_{1-x})_2O$ similar to Eq. (2):

$$2(1-x)Ga(a) + In_2O_3(a) \rightarrow (In_xGa_{1-x})_2O_3(s) + 2(1-x)In(a) \quad (7)$$

Direct formation of $(In_xGa_{1-x})_2O_3$ from adsorbed Ga, In, and O is inhibited by the preferential oxidation of In.[53] Cation exchange between Ga and $In_2O_3$ is therefore expected to be the primary reaction pathway for growth of $(In_xGa_{1-x})_2O_3$.

Excess adsorbed Ga may decompose the $(In_xGa_{1-x})_2O_3$ film in a modified version of reactions (2) and (3):[54]

$$2(r_{Ga}-(1-x))Ga(a) + (In_xGa_{1-x})_2O_3(s) \rightarrow 2xIn(a) + Ga_2O_3(s) + 2(r_{Ga}-1)Ga(a), \quad (8a)$$

$$Ga_2O_3(s) + 2(r_{Ga}-1)Ga(a) \rightarrow \tfrac{3}{2}(r_{Ga}-1)Ga_2O(a) + \tfrac{1}{2}(3-r_{Ga})Ga_2O_3(s) \quad (8b)$$

Where $r_{Ga} = \Phi_{Ga}/\Phi_O^{*,Ga}$. The excess Ga first displaces In from $(In_xGa_{1-x})_2O_3$ to form $Ga_2O_3$ and adsorbed In, and subsequently decomposes $Ga_2O_3$ to form the suboxide $Ga_2O$. The adsorbed In and $Ga_2O$ readily desorb at the growth temperatures investigated,[52,55] resulting in only a $Ga_2O_3$ film remaining.

At sufficiently high Ga flux and growth temperatures above ~725 °C, Ga exchange with In occurs according to reactions (2) and (6), consuming the $In_2O_3$ and forming $(In_xGa_{1-x})_2O_3$ or $Ga_2O_3$. However, this cation exchange process is inhibited at low growth temperatures[54] and low Ga fluxes, resulting in the accumulation of $In_2O_3$ on the growth surface and the appearance of the spotty/faceted surface reconstruction. Additionally, increasing the In flux at high growth temperatures does not result in the accumulation of $In_2O_3$ as the excess In flux instead forms volatile $In_2O$ according to reaction (5). The spotty/faceted RHEED pattern associated with bixbyite $In_2O_3$ is therefore only stabilized under relatively low growth temperature, low metal flux conditions where $In_2O_3$ is not consumed by either Ga cation exchange or by $In_2O$ suboxide formation.

### 3. *In-plane lattice spacing analysis from RHEED*

In the cyclical MBE methodology, sample characterization may be accomplished using techniques compatible with ultra-high vacuum growth, such as spectroscopic ellipsometry, mass spectroscopy, or RHEED. In particular, RHEED is a highly surface-sensitive technique that probes the first few atomic layers of the sample[36,37] and is capable of monitoring crystal growth in real time. RHEED provides information about surface chemistry and reconstruction, lattice periodicity, surface roughness, crystalline phases and domain size, growth rate, and growth mode (e.g. layer-by-layer vs step flow) for thin films.[36,37,56,57] The ability to measure lattice periodicity, corresponding to crystal phase and composition, and diffraction pattern FWHM, corresponding to surface roughness and domain size, in a real-time manner make RHEED uniquely suited to the study of cyclical growth and etch of β-$Ga_2O_3$ based alloys.

RHEED offers a wealth of information about crystal growth. The *d*-spacing of the first-order streaks directly measures the lattice periodicity of the uppermost atomic layers of the growing crystal. The FWHM of the (00) streak measures the crystal domain size and is correlated with surface roughness. The periodicity of the surface reconstruction, e.g. 2×, is directly related to the chemical bonds and crystal phase present at the sample surface. The appearance of bright diffraction spots indicates beam transmission through a roughened surface of 3D islands.[36,57] Finally, the intensity of the specular reflection and its evolution with time provide information about growth mode.[37,56] For example, layer-by-layer (Frank-van der Merwe) growth results in oscillations in RHEED intensity. The period of this oscillation is the time to form a complete monolayer. In contrast, step-flow growth and island growth does not result in intensity oscillations. By tracking the evolution of these RHEED features over time, a nearly complete picture of crystal growth is deduced.

RHEED was used to analyze a change in the in-plane lattice spacing between the principal RHEED streaks during the $(In_xGa_{1-x})_2O_3$ growth in many growth conditions. The percentage change in *d*-spacing between the $(01)$ and $(0\bar{1})$ streaks is shown for an In flux dependent growth series in Fig. 5. The percentage change in *d*-spacing is normalized to the β-$Ga_2O_3$ *c*-plane lattice constant according to $[(2d - c)/c] \cdot 100\%$. The *d*-spacing does not change with time for homoepitaxial growth of $Ga_2O_3$ at 723 °C and $1.5\times10^{-7}$ torr Ga flux. The *d*-spacing increases by as much as 1.2% immediately upon initiation of $(In_xGa_{1-x})_2O_3$ growth with In flux ranging from $1\times10^{-7}$ to $3\times10^{-7}$ torr. The *d*-spacing increases slowly for In flux of 3-$4\times10^{-7}$ torr and contracts slightly for high In fluxes of $5\times10^{-7}$ torr and greater. Expansion of the *d*-spacing is associated with the growth of an In-containing film with larger unit cell than the β-$Ga_2O_3$ substrate, since the crystal termination layer is free to reconstruct and may relax with respect to the substrate lattice constants, even if the underlying bulk crystal is fully strained. The rapid increase in *d*-spacing at the beginning of growth, and subsequent slow decrease, suggest that lower In fluxes of $1\times10^{-7}$ to $3\times10^{-7}$ torr result in growth of initially In-rich films with decreasing In incorporation over time. Vegard's Law interpolation of the β-$Ga_2O_3$ and monoclinic $In_2O_3$ *c*-plane lattice constants indicates approximately 10% In mole fraction for a *d*-spacing increase of 1%.[31,33] By contrast, growth at high In fluxes does not result in a significant change in *d*-spacing, suggesting negligible In incorporation.

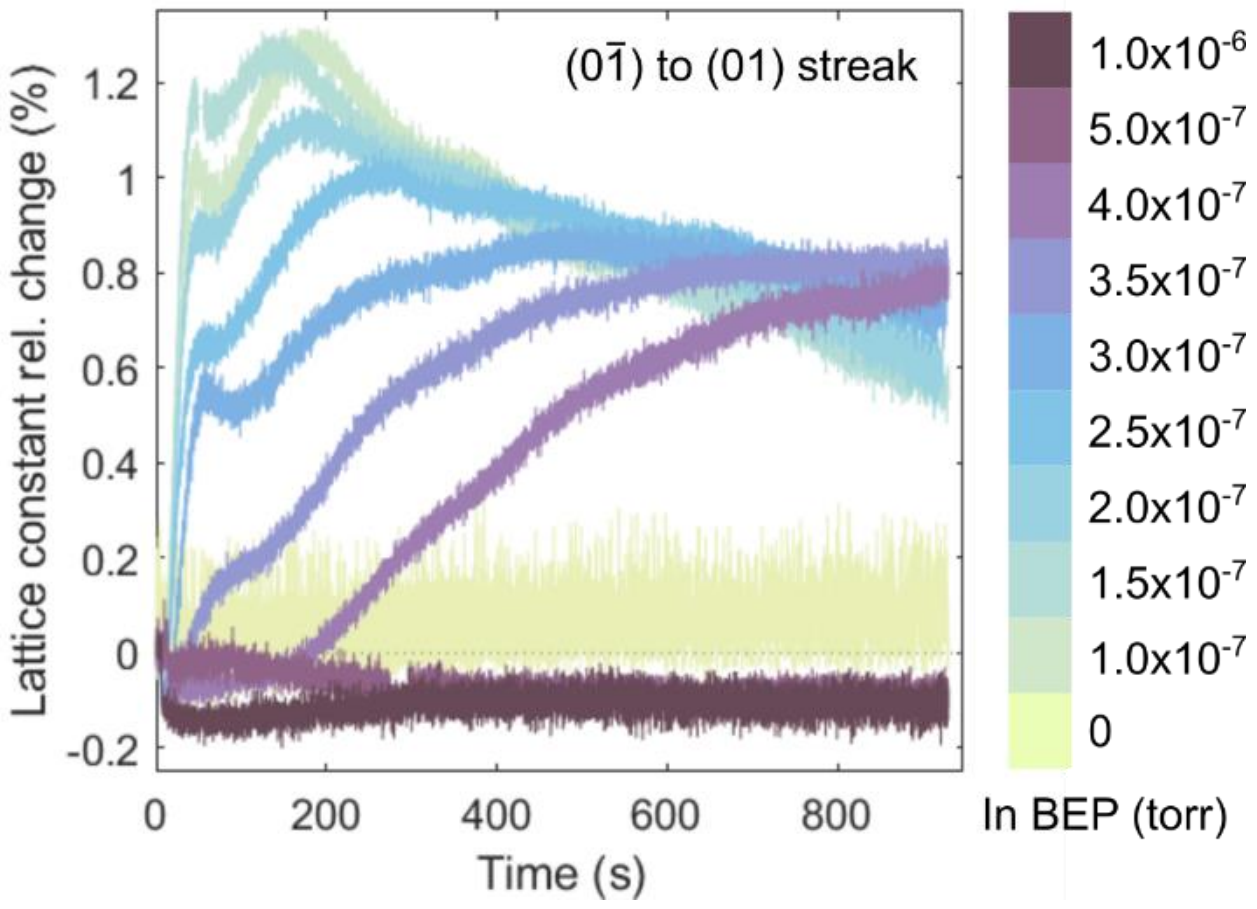

FIG. 5. Percentage change in *d*-spacing relative to the $Ga_2O_3$ *c* plane lattice constant vs growth time for $(0\bar{1})$ to $(01)$ streak spacing. All $(InGa)_2O_3$ growths were performed at 723 °C and Ga BEP = $1.5\times10^{-7}$ torr. In BEP in torr is indicated by color bar and varies from zero to $1\times10^{-6}$ torr.

This counter-intuitive trend of decreasing In incorporation vs increasing In flux can be explained by the reaction equations (4) through (6). At low In fluxes corresponding to the oxygen-rich regime (Eq. 4), $In_2O_3$ is readily formed and cation exchange with Ga occurs to form $(In_xGa_{1-x})_2O_3$. The formation of $(In_xGa_{1-x})_2O_3$ subtracts from the amount of $In_2O_3$ present on the growth surface, leading to a gradual reduction in the growth rate and In content of $(In_xGa_{1-x})_2O_3$ over time. This is evident in the drop in d-spacing vs time for In fluxes ranging from 1-$2\times10^{-7}$ torr. As the In flux increases to the moderately In-rich regime, $In_2O$ suboxide formation reduces the amount of $In_2O_3$ available for Ga cation exchange, evident in the markedly smaller initial d-spacing shift for In fluxes ranging from 2.5-$4\times10^{-7}$ torr. In the highly In-rich regime (Eq. 5), no $In_2O_3$ is formed and all available oxygen flux is consumed as $In_2O$. Therefore either no film growth occurs, or only a homoepitaxial $Ga_2O_3$ film with negligible In incorporation is grown, and the d-spacing remains constant with time. Under highly In-rich conditions with flux greater than $5\times10^{-7}$ torr, Ga suboxide etching is again expected to occur since no adsorbed oxygen is present and the adsorbed Ga flux can react only with the $Ga_2O_3$ substrate.

### 4. *Ex-situ analysis of targeted $(In_xGa_{1-x})_2O_3$ growth*

A targeted growth was conducted based on the RHEED surface reconstruction map in Fig. 4. Growth conditions were selected within the streaky 2× region and near the boundary with spotty/faceted $In_2O_3$ growth, and $(In_xGa_{1-x})_2O_3$ was grown for 4 hours. These conditions were expected to yield significant In incorporation in monoclinic $(In_xGa_{1-x})_2O_3$.

Fig. 6a shows a XRD reciprocal space map (RSM) about the β-$Ga_2O_3$ (420) reflection for the 4 hour long targeted growth performed at 750 °C, $1.5\times10^{-7}$ torr Ga flux, and $2.0\times10^{-7}$ torr In flux. A high resolution coupled ω-2θ scan about the (020) reflection is shown inset. Significant In incorporation is evident from the compressively strained features at low reciprocal space coordinate $q_z$. The film is coherently strained to the β-$Ga_2O_3$ substrate with only minimal relaxation to lower reciprocal space coordinate $q_x$. The RSM confirms that $(In_xGa_{1-x})_2O_3$ alloy growth is the monoclinic *C2/m* phase and phase separation does not occur. The absence of a single well-defined compressive peak in the coupled ω-2θ scan indicates that In incorporation is highly non-uniform in the growth direction.

Fig. 6b shows the XPS and simulated RBS atomic concentration depth profiles for the targeted growth. The indium concentration curves are scaled by a factor of 10 for clarity. The atomic concentrations at the sample surface are reduced due to the presence of a thin Au layer deposited to prevent sample charging during RBS measurements. Peak In mole fraction of $x$ = 5.6% in $(In_xGa_{1-x})_2O_3$ is observed. The In mole fraction decreases in the growth direction, in qualitative agreement with the RHEED *d*-spacing decrease over time observed in Fig. 5. The simulated film thickness is approximately 980 nm. Additional targeted growths conducted at higher Ga fluxes (not shown) did not yield In incorporation measurable by XPS or RBS. This illustrates the competition between In-incorporating $(In_xGa_{1-x})_2O_3$ versus In-catalyzed $Ga_2O_3$ growth regimes. In incorporation is achieved at reduced Ga fluxes, in agreement with previous studies[54,58] which indicated full In incorporation at very low metal/oxygen ratios (highly oxygen rich growth). The In-catalyzed growth regime at higher Ga fluxes achieves high growth rates, however In incorporation is suppressed.

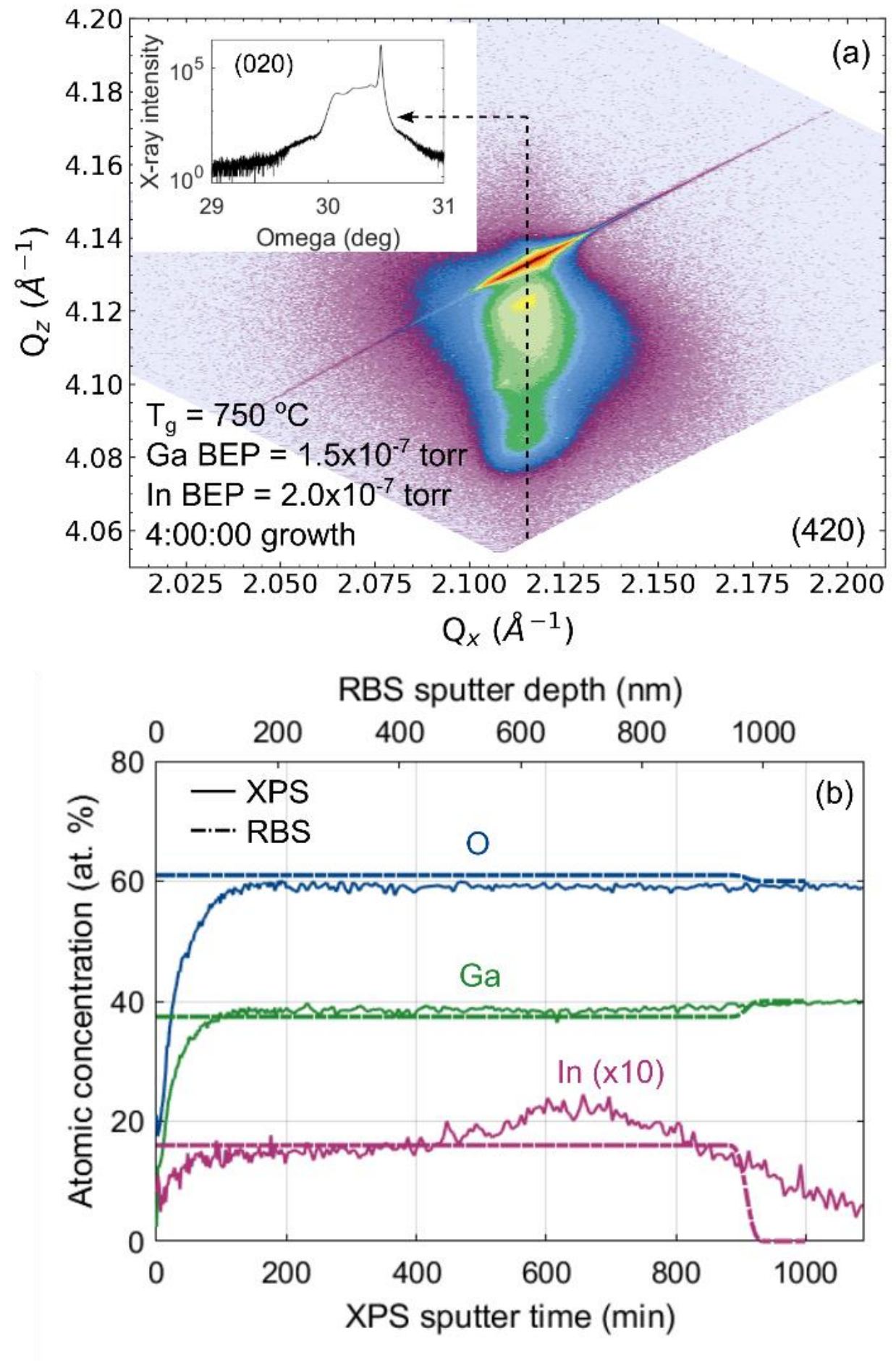

FIG. 6. Ex-situ characterization of high-In content targeted $(InGa)_2O_3$ growth. (a) Reciprocal space map of (420) plane. Growth conditions are labeled directly on figure. Inset: Coupled ω-2θ scan of (020) plane. (b) Depth profiles of atomic concentration of Ga, In, and O for high-In content targeted thick $(In_xGa_{1-x})_2O_3$ growth. XPS measurement is shown by solid lines. The simulated depth profile for RBS measurements is shown as dot-dashed curves. Indium concentration curves are scaled by a factor of 10 for clarity.

## I. SUMMARY AND CONCLUSIONS

In summary, the results of the cyclical growth/etch RHEED experiments and the targeted thick growth paint a complete picture of the growth space for MBE grown monoclinic $(In_xGa_{1-x})_2O_3$ alloys. Growth proceeds as a two-step process characterized by the preferential oxidation of adsorbed In followed by cation exchange with adsorbed Ga. At low growth temperatures below approximately 725 °C, cation exchange is inhibited and bixbyite $In_2O_3$ forms. At high growth temperatures, cation exchange occurs and $(In_xGa_{1-x})_2O_3$ growth is dominated by the effective metal/oxygen flux ratios. Low metal flux ratios, i.e. oxygen rich growth, yield maximum In incorporation at the expense of low growth rate. Counter-intuitively, moderately In-rich growth with $1 < r_{In} < 3$ results in a *reduction* of In incorporation and growth rate due to the formation of $In_2O$ suboxide. Very In-rich growth with $r_{In} >$ 3 results in no film growth, or only very slow growth of bixbyite $In_2O_3$, due to nearly complete consumption of all available oxygen flux as $In_2O$. Increasing $r_{Ga}$ at low to moderate $r_{In}$ results in reduced In incorporation but increased $(In_xGa_{1-x})_2O_3$ growth rate due to Ga cation exchange. As $r_{Ga}$ is increased further to moderately Ga-rich conditions, $Ga_2O$ suboxide formation decomposes the $(In_xGa_{1-x})_2O_3$ film and reduces both In incorporation and growth rate further. Adsorbed In displaced from the $(In_xGa_{1-x})_2O_3$ film readily desorbs at these growth temperatures, with the result that the growth surface maintains a streaky 2× surface reconstruction typical of $Ga_2O_3$ homoepitaxial growth. MBE growth at conditions near the RHEED surface reconstruction boundary between streaky 2× and spotty/faceted patterns is therefore expected to yield maximum In incorporation at a given $(In_xGa_{1-x})_2O_3$ growth rate determined by the Ga flux. The RHEED streak *d*-spacing provides a measure of the epilayer lattice constant, and its evolution vs time illustrates the non-uniform nature of In incorporation in $(In_xGa_{1-x})_2O_3$. This non-uniform composition is confirmed by *ex-situ* XRD and RBS measurements of a targeted growth which exhibits peak In mole fraction of 5.6% in a coherently strained monoclinic $(In_xGa_{1-x})_2O_3$ film. Future work could focus on optimization of In incorporation uniformity and material quality by real-time control of Ga and In flux and/or substrate temperature.

The cyclical oxide growth and etch-back method developed in this work stands apart from traditional MBE growth techniques which rely on unidirectional material deposition. Leveraging the unique sub-oxide desorption under metal flux and a differentially pumped RHEED system, a six-fold improvement in experimental throughput was achieved. The growth and etch-back process is remarkably repeatable, with full recovery of the β-$Ga_2O_3$ substrate growth surface roughness (~1 nm) achieved up to 46 cycles on a single wafer. Indefinite substrate re-use can be achieved with an In-free mounting scheme. RHEED images of surface reconstruction, line scans of streak *d*-spacing and FWHM vs time, and streak intensity profiles provide both qualitative and quantitative feedback to the grower in real-time. This cyclical growth and etch-back method can be applied to other material systems exhibiting incongruent subcomponent evaporation from a two-phase mixture, including both oxides (e.g. $GeO_2$, $SnO_2$) and other chemistries (e.g. $In_2Se_3$, InSe). In addition to the examples of $Ga_2O_3$ and $In_2O_3$ investigated here, this class of oxide materials includes numerous rare earth-oxides, lanthanide-oxides, and the wide-bandgap group IV-oxides $SiO_2$, $SnO_2$, and $GeO_2$ which evaporate as $IV(a) + IVO_2(s) \rightarrow 2IVO(g)$.[35] $GeO_2$ is a material of particular interest as it has an ultra wide bandgap of 4.6-4.7 eV[59] and ambipolar dopability (i.e. *n* and *p* type conductivity).[60] $GeO_2$ remains elusive to growth by MBE[61] and mist CVD.[62] Further refinements in RHEED image analysis will complement this inherently high-throughput experimental methodology to rapidly characterize the growth space for both $Ga_2O_3$ based alloys and other emerging materials.

## II. CONFLICTS OF INTEREST

There are no conflicts of interest to declare.

## III. ACKNOWLEDGEMENTS

This work was authored by the National Renewable Energy Laboratory (NREL), operated by Alliance for Sustainable Energy, LLC, for the US Department of Energy (DOE) under Contract No. DE-AC36-08GO28308. Funding provided by the Laboratory Directed Research and Development (LDRD) Program at NREL, and by the Office of Energy Efficiency and Renewable Energy (EERE), Advanced Manufacturing Office. The views expressed in the article do not necessarily represent the views of the DOE or the US Government.

## IV. AUTHOR CONTRIBUTIONS

**Stephen Schaefer**: Conceptualization, data curation, formal analysis, investigation, methodology, visualization, writing (original draft)

**Davi Febba**: Formal analysis, data curation, investigation, methodology, software, validation, writing (original draft)

**Kingsley Egbo**: Investigation, methodology, resources, writing (editing and review)

**Glenn Teeter**: Investigation, methodology, resources

**Andriy Zakutayev**: Conceptualization, funding acquisition, supervision, writing (editing and review)

**Brooks Tellekamp**: Conceptualization, formal analysis, funding acquisition, investigation, methodology, project administration, resources, supervision, writing (editing and review)

## SUPPLEMENTAL INFORMATION

### 1. *Flux calculations*

The Ga and In beam equivalent pressures are converted into particle fluxes $\Phi_{Me}$ ($cm^{-2}s^{-1}$) in Fig. 4b-d according to kinetic gas theory:[58]

$$\Phi_{Me}^{BEP} = C\frac{\rho_{Me}}{\sqrt{2\pi m_{Me} k_B T}}, \quad (9)$$

$$\Phi_{Me} = I_{Me}\Phi_{Me}^{BEP}, \quad (10)$$

Where $m_{Me}$ is the atomic mass, $T$ is the effusion cell temperature, $C$ = 0.013332 $cm^{-2}s^{-1}$ is a prefactor to convert $\Phi_{Me}^{BEP}$ from torr to particle flux, and $I_{Me}$ is an effective sensitivity factor determined from metal-limited calibration growth. The Ga sensitivity factor $I_{Ga}$ was determined from homoepitaxial growth of $Ga_2O_3$ with a typical growth rate of 3.16 nm/min, and the In sensitivity factor $I_{In}$ was estimated as $I_{In} = I_{Ga}(5.2/9.1)$ in accordance with the experimentally obtained factors reported in Ref. 60. The oxygen beam equivalent pressure $\Phi_O^{BEP}$ = $3.2\times10^{15}$ $cm^{-2}s^{-1}$ at 3.00 SCCM and 250 W was calculated from the growth rate of slightly metal-rich homoepitaxial $Ga_2O_3$ growth at 650 °C. The effective oxygen flux $\Phi_O^{*,Me}$ available for $Ga_2O_3$ and $In_2O_3$ growth varies for each metal species due to the differing oxidation efficiencies[34,54,58] and is given by:

$$\Phi_O^{*,Me} = J_O^{Me}\Phi_O^{BEP}. \quad (11)$$

In the presence of both Ga and In flux, the total effective oxygen flux $\Phi_O^*$ is modified by the metal flux ratio $R = \Phi_{In}/(\Phi_{Ga} + \Phi_{In})$, with[58]

$$\Phi_O^* = \Phi_O^{BEP}(RJ_O^{In} + (1-R)J_O^{Ga}). \quad (12)$$

The total metal/oxygen flux ratio is thus $(\Phi_{Ga} + \Phi_{In})/\Phi_O^*$ and the stoichiometric flux ratio shown in Fig. 4d is $\frac{3}{2}(\Phi_{Ga} + \Phi_{In})/\Phi_O^*$. The flux calibration parameters are summarized in Table I.

TABLE I. Flux calibration parameters for Ga and In.

| Metal | Ga | In |
|---|---|---|
| Atomic mass, $m_{Me}$ (amu) | 69.723 | 114.818 |
| Sensitivity factor, $I_{Me}$ | 10.60 | 6.06 |
| Oxidation efficiency, $J_O^{Me}$ | 0.096 | 0.263 |

### 2. *X-ray diffraction of phase separated growth*

Fig. 7 shows a coupled ω-2θ x-ray diffraction scan for a targeted growth performed at 740 °C with relatively high In and Ga fluxes of $3\times10^{-7}$ torr each. The dashed lines indicate the locations of the β-$Ga_2O_3$ (020) reflection at ω = 30.481° and the bixbyite $In_2O_3$ (444) reflection at ω = 31.824°. The presence of the measured bixbyite $In_2O_3$ (444) peak reflects the terminal RHEED pattern which developed into a spotty/faceted pattern typical of Fig. 3b, confirming that this pattern is due to the formation of $In_2O_3$ on the surface. No compressive features are observed at diffraction angles below the β-$Ga_2O_3$ (020) peak, confirming that In does not incorporate at these high metal flux growth conditions. RBS measurements for this sample (not shown) confirm the presence of a thin In-rich layer at the surface. Reaction equations (4) through (8) suggest that $In_2O_3$ formation is suppressed at these metal rich conditions due to a combination of $In_2O$ formation and Ga cation exchange followed by $Ga_2O$ decomposition. The presence of a thin $In_2O_3$ layer suggests that complete decomposition of $In_2O_3$ does not occur. The accumulation of $In_2O_3$ is very small, however, on the order of several nanometers over the course of a 1 hour growth.

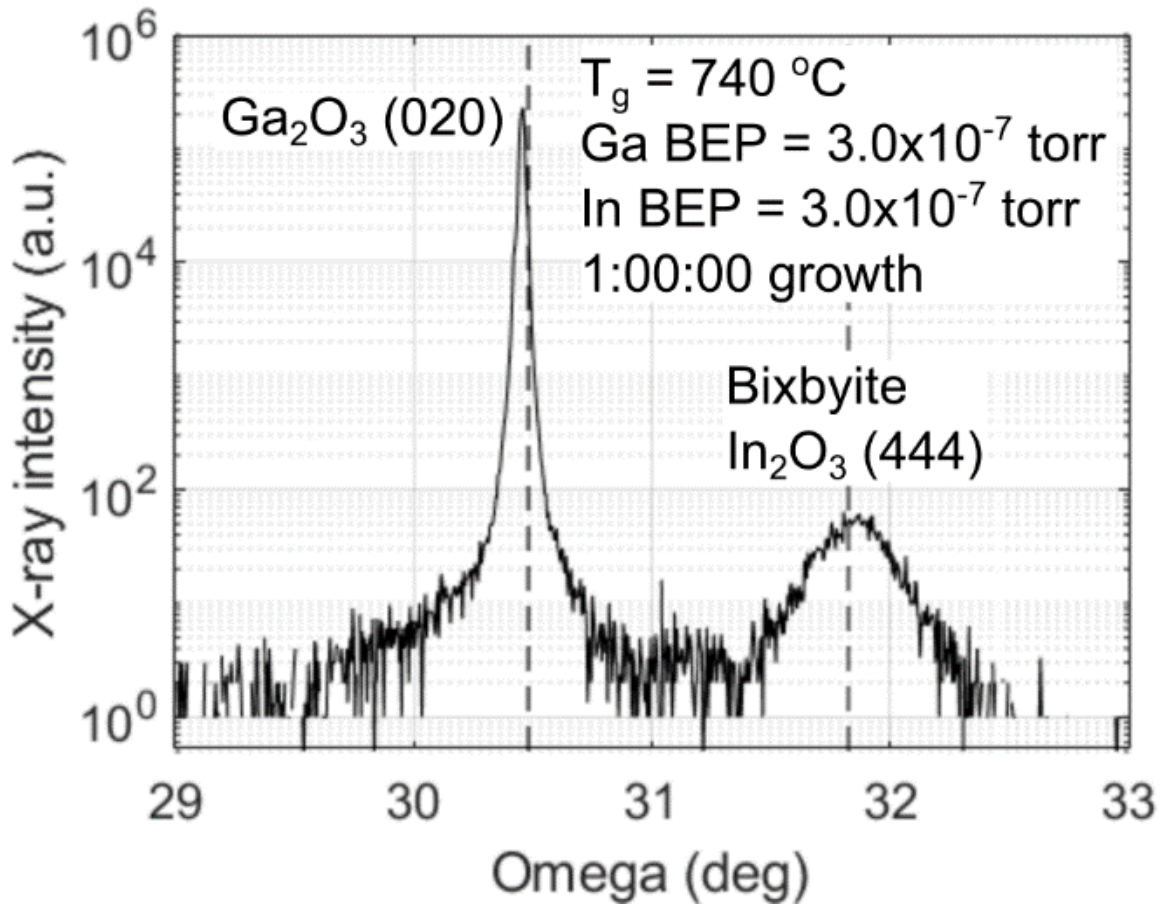

FIG. 7. Coupled ω-2θ x-ray diffraction (XRD) scan of targeted growth with spotty/faceted surface reconstruction. Growth conditions labeled directly on figure. The peak at ω = 31.824° is attributed to the (444) reflection from bixbyite $In_2O_3$, responsible for the spotty/faceted RHEED pattern.

### 3. *Orientation of bixbyite $In_2O_3$ on (010) β-$Ga_2O_3$*

Fig. 8 shows the possible orientation of bixbyite $In_2O_3$ on (010) oriented β-$Ga_2O_3$. The $In_2O_3$ [111] direction is parallel to the $Ga_2O_3$ [010] direction and is perpendicular to the growth plane of the substrate. The $In_2O_3$ [011] direction is perpendicular to the $Ga_2O_3$ [001] direction. This orientation is responsible for the bixbyite $In_2O_3$ (444) peak observed in the coupled ω-2θ XRD scan shown in Fig. 7 for the phase separated growth, and for the ~7.1 Å spot spacing observed in the RHEED pattern shown in Fig. 3b corresponding to diffraction from the $In_2O_3$ (110) plane.

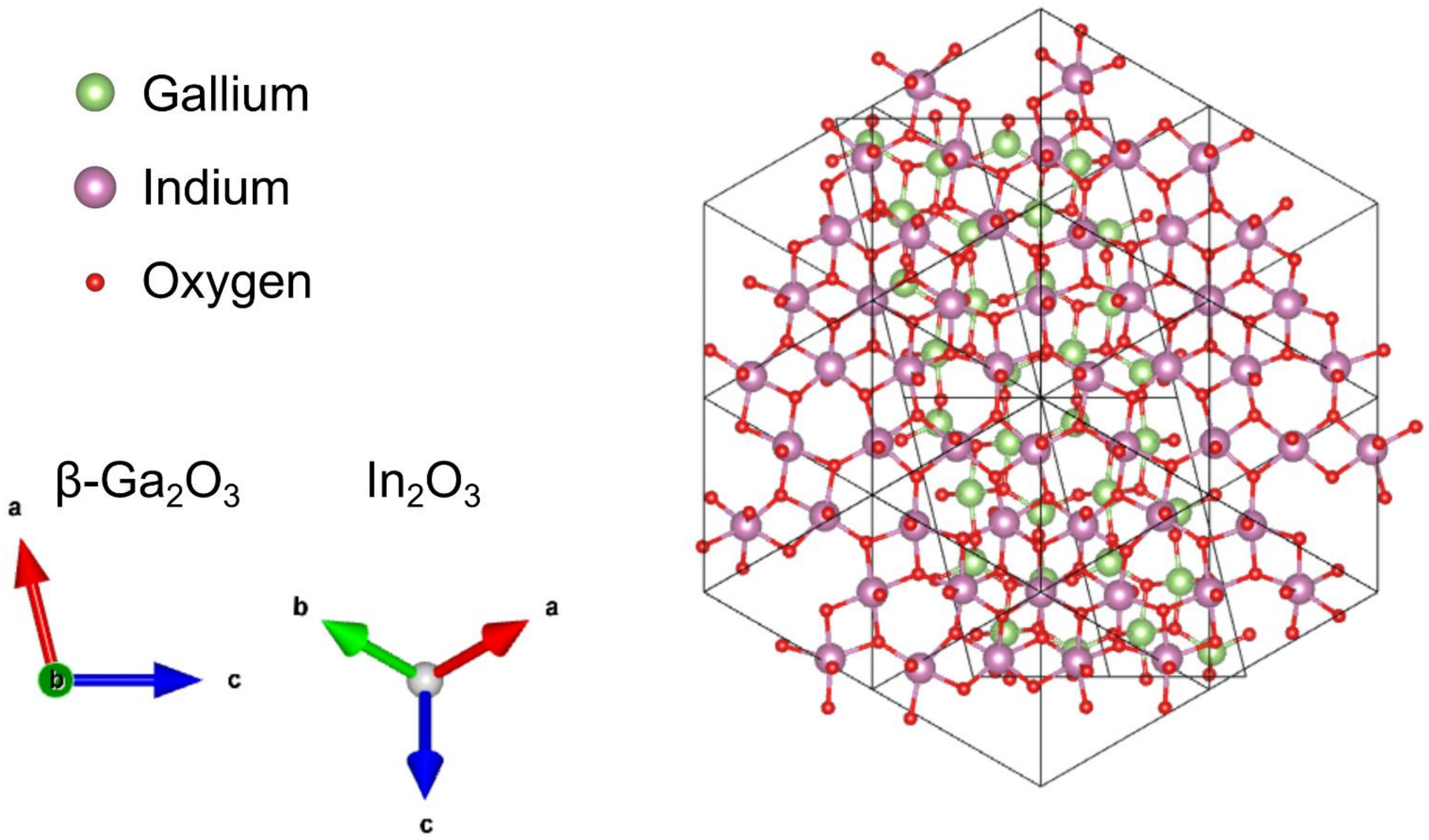

FIG. 8. Possible orientation of bixbyite $In_2O_3$ on (010) oriented β-$Ga_2O_3$.